\documentclass[a4paper,conference]{IEEEtran}
\IEEEoverridecommandlockouts
\usepackage{cite}
\usepackage{amsmath,amssymb,amsfonts}
\usepackage{algorithmic}
\usepackage{adjustbox}
\usepackage{algorithm}
\usepackage{graphicx}
\usepackage{multirow}
\usepackage{mfirstuc}
\usepackage{textcomp}
\usepackage{xcolor}
\usepackage{amssymb}
\usepackage{svg}
\usepackage{amsmath}
\usepackage{caption}
\usepackage{subcaption}
\usepackage{amsfonts}
\usepackage{url}
\usepackage{color}
\usepackage{ragged2e}
\definecolor{blue}{rgb}{0,0,1}

\mathchardef\mhyphen="2D
\usepackage{lineno,hyperref}
\usepackage{soul}
\usepackage{nomencl}
\makenomenclature

\begin{document}
	\title{A Multi-Objective Planning and Scheduling Framework for Community Energy Storage Systems in Low Voltage Distribution Networks\\
	}

\makeatletter
\newcommand{\linebreakand}{%
  \end{@IEEEauthorhalign}
  \hfill\mbox{}\par
  \mbox{}\hfill\begin{@IEEEauthorhalign}
}
\makeatother

\author{\IEEEauthorblockA{K.B.J. Anuradha}
		\vspace{0.4mm}
		\IEEEauthorblockN{
			\textit{The Australian National University}\\
			\vspace{0.4mm}
			Canberra, Australia\\
			\vspace{0.4mm}
			Jayaminda.KariyawasamBovithanthri@anu.edu.au}
		\vspace{0.4mm}
		\and
		\IEEEauthorblockN{Chathurika P. Mediwaththe}
		\vspace{0.4mm}
		\IEEEauthorblockA{
			\textit{The Australian National University \& CSIRO}\\
			\vspace{0.4mm}
			Canberra, Australia\\
			\vspace{0.4mm}
			chathurika.mediwaththe@csiro.au}
		
		\linebreakand
		\IEEEauthorblockN{Masoume Mahmoodi}
		\vspace{0.3mm}
		\IEEEauthorblockA{
			\textit{The Australian National University}\\
			\vspace{0.4mm}
			Canberra, Australia\\
			\vspace{0.4mm}
			masoume.mahmoodi@anu.edu.au}
}
	
\setlength{\nomitemsep}{1.1mm}	
\maketitle
\renewcommand\nomgroup[1]{%
  \item[
  \ifstrequal{#1}{A}{\vspace{50mm} \textit{Sets and Indices}}{%
  \ifstrequal{#1}{C}{\textit{Model Parameters}}{%
  \ifstrequal{#1}{B}{\textit{Functions}}{%
  \ifstrequal{#1}{E}{\textit{Planning Decision Variables}}{%
  \ifstrequal{#1}{F}{\textit{Operation Decision Variables}}{%
  \ifstrequal{#1}{Z}{\textit{Other Notations}}{%
  \ifstrequal{#1}{D}{\textit{Power Flows and Injections}}{}}}}}}}%
]}

\makenomenclature
\nomenclature[A, 01]{\(\mathcal V\), \(i\), \(j\)}{ Set of nodes, node indices}
\nomenclature[A, 02]{\(\mathcal E\)}{ Set of lines in the network}
\nomenclature[A, 03]{\(\mathcal W_{j}\)}{  Set of downstream nodes of node j including itself}
\nomenclature[A, 04]{\(\mathcal C_{j}\), \(c\)}{  Set of customers at node $j$, customer index}
\nomenclature[A, 05]{\(\mathcal T\), \(t\)}{ Set of time intervals, time index}
\nomenclature[A, 06]{$X$, \textbf{x}}{ Feasible set, decision variable vector}

\nomenclature[C, 01]{\( r_{ij},  x_{ij}\)}{ Resistance and reactance of line $(i,j)$ - $(\Omega) $}
\nomenclature[C, 02]{$U_{min}$, $U_{max}$}{ Minimum and maximum squared voltage magnitude limits - $(V^2)$}
\nomenclature[C, 03]{$\eta^{ch}$, $\eta^{dis}$}{ Charging and discharging efficiencies of the CES}
\nomenclature[C, 04]{$\lambda _{min}$, $\lambda _{max}$}{ Percentage coefficients of the CES capacity }
\nomenclature[C, 05]{$a_{j}$}{ Binary variable to find the optimal CES location}
\nomenclature[C, 06]{$p_{j}^{Rate}$}{ Optimal CES rated power- $(kW)$  }
\nomenclature[C, 07]{$E_{j}^{cap}$}{ Optimal CES capacity - $(kWh)$  }
\nomenclature[C, 08]{$p_{min}^{Rate}$, $p_{max}^{Rate}$}{Minimum and maximum allowable CES rated power- $(kW)$ }
\nomenclature[C, 09]{$E_{min}^{CES}$, $E_{max}^{CES}$}{Minimum and maximum allowable CES capacity - $(kWh)$ }
\nomenclature[C, 10]{$\lambda _{p}(t)$}{Grid energy price at time $t$ - $(AUD/kWh)$ }
\nomenclature[C, 11]{$\gamma_{CES}$}{Fixed part of the CES investment cost - $(AUD)$ }
\nomenclature[C, 12]{$\delta_{CES}$}{ Cost of the CES for a unit capacity - $(AUD/kWh)$ }
\nomenclature[C, 13]{$\Delta t$}{ Time difference between two adjacent time instances - $(h)$ }
\nomenclature[C, 14]{$w_{i}$}{ Weight coefficient of the $i^{th}$ objective function }

\nomenclature[D, 01]{\( p_{cj}^{L}(t), q_{cj}^{L}(t)\)}{ Real and reactive power consumption of the customer $c$ at node $j$ at time $t$ - $(kW, kVAR)$  }
\nomenclature[D, 02]{\( p_{cj}^{PV}(t) \)}{ SPV generation of the customer $c$ at node $j$ at time $t$ - $(kW)$  }
\nomenclature[D, 03]{\( P_{ij}(t), Q_{ij}(t)\)}{ Real and reactive power flow from $i$ to $j$ node at time $t$ - $(kW, kVAR)$  }
\nomenclature[D, 04]{\( p_{j}(t), q_{j}(t)\)}{ Real and reactive power absorption at node $j$ at time $t$ - $(kW, kVAR)$  }
\nomenclature[D, 05]{\(  p_{cj}^{G}(t))\)}{ Real power exchange with the grid by the customer $c$ at node $j$ at time $t$ - $(kW)$  }
\nomenclature[D, 06]{\(  p_{cj}^{CES}(t))\)}{ Real power exchange with the CES by the customer $c$ at node $j$ at time $t$ - $(kW)$ }
\nomenclature[D, 07]{\(  p_{CES}^{G}(t))\)}{ Real power exchange with the grid by the CES at time $t$- $(kW)$  }
\nomenclature[D, 08]{\( p_{j}^{CES,ch}(t)\)}{ Charging power of the CES at node $j$ at time $t$ - $(kW)$  }
\nomenclature[D, 09]{\( p_{j}^{CES,dis}(t)\)}{ Discharging power of the CES at node $j$ at time $t$ - $(kW)$ }  

\nomenclature[Z, 01]{$E_{j}^{CES}(t)$}{ Energy level of the CES at node $j$ at time $t$  - $(kWh)$  }
\nomenclature[Z, 02]{$V_{j}(t)$}{ Voltage magnitude of node  $j$ at time $t$ - $(V)$}
\nomenclature[Z, 03]{$U_{j}(t)$}{ Squared voltage magnitude of node $j$ at time $t$ - $(V^2)$}
\nomenclature[Z, 04]{$I_{ij}(t)$}{ Current flow from node $i$ to $j$ at time $t$ - $(A)$}

 \begin{abstract}
This paper presents a methodology for optimizing the planning and scheduling aspects of a community energy storage (CES) system in the presence of solar photovoltaic (SPV) power in low voltage (LV) distribution networks. To this end, we develop a multi-objective optimization framework that minimizes the real power loss, the energy trading cost of LV customers and the CES provider with the grid, and the investment cost for the CES. Distribution network limits including the voltage constraint are also taken into account by combining the optimization problem with a linearized power flow model. Simulations for the proposed optimization framework with real power consumption and SPV generation data of the customers, highlight both real power loss and energy trading cost with the grid are reduced compared with the case without a CES by nearly 29\% and 16\%, respectively. Moreover, a case study justifies our methodology is competent in attaining the three objectives better than the optimization models which optimize only the CES scheduling.
		
\end{abstract}
	
	\vspace{1.5mm}
        \renewcommand\IEEEkeywordsname{Keywords}
	
\printnomenclature[2.5cm]	
 
\section{{\normalsize Introduction}}
	
In the recent past, there has been a notable interest among the power systems research community and the industry for the uptake of community energy storage (CES) in low voltage (LV) power systems. This trend is driven by the benefits gained from a CES such as providing the opportunity to increase the hosting capacity of the network, enhancing the solar energy self consumption of the customers, and increasing the community access to renewable energy \cite{Marnie2020Community}. Additionally, CES devices can be deployed to gain technical merits such as real power loss minimization and economic benefits including the curtailment of energy purchase cost of the customers \cite{Zheng2020Hierarchical}.
	
As discussed in literature, a CES  may be used in energy management problems to earn technical and monetary benefits together \cite{Chathurika2021Network,Chathurika2020Community}. Those merits can be fully exploited if the CES planning aspects including its location, the rated power and the capacity are optimized simultaneously with the CES scheduling aspects namely, its charging and discharging. 
	
The existing literature on CES utilization in LV distribution networks can be divided into two categories as; (i) optimization of CES scheduling only, (ii) optimization of both CES planning and scheduling. In the first category, the authors have presented optimization frameworks for  CES scheduling without accounting for its planning aspects. For instance, a method built up on game theory concepts to maximize the revenue for the CES provider and minimize energy costs for the customers is discussed in  \cite{Chathurika2021Network}. A multi-objective framework to minimize the real energy loss and energy costs of the customers and the CES provider for trading energy with the grid is discussed in \cite{Chathurika2020Community}. A method based on model predictive control to optimize the CES scheduling is presented in \cite{Multi2018Zafar}. 
	
In addition to the papers which have presented methods for optimizing only the CES scheduling, there are research work which have proposed methods for optimizing both planning and scheduling of CES simultaneously. For instance, a method for maximizing the hosting capacity in a distribution network in the presence of a CES is proposed in \cite{HasanporDivshali2019Hosting}. Analytical methods to minimize the real energy loss of a network by finding the optimal CES location and its capacity are discussed in \cite{Hung2011Community,Bohringer2021Sizing}. A common feature of these methods is that the optimal CES planning
aspects are determined based on analytical (such as graphical or
numerical methods) and sensitivity based approaches (methods
which decide the optimal values based on a calculated sensitivity
parameter). These approaches can
be computationally exhaustive as the optimal CES location and the
capacity are found upon computing a sensitivity parameter
for a large number of location-capacity combinations. Also,
even after an exhaustive search, it is not always guaranteed
to reach an optimal solution \cite{Hung2011Community}. Thus, a robust formulation to optimize the capacity, the rated power and the location of a CES while generating the techno-economic benefits associated with such storage devices would be an effective alternative to overcome the challenges in the literature.

In this paper, we study the extent to which the location,
the capacity and the power rating of a CES in addition to its
scheduling, affect network and economic benefits achievable
from it. For this, we develop an optimization framework that
optimizes both planning and scheduling of a CES. The optimized
planning and scheduling aspects are then leveraged to
minimize the network power loss, cost incurred by the customers and the
CES provider for trading energy with the grid and the investment
cost of the CES simultaneously. To the best of our knowledge,
this problem has not been addressed in the literature. The contributions of this paper are as follows.

\begin{itemize}
    \item A linearized power flow model is exploited with the CES operational constraints to develop a multi-objective optimization framework. It is then solved as a mixed integer quadratic program according to the optimization algorithms in \cite{Boyd2021Optimization}. The analytic hierarchy process (AHP) is used for fairly weighting the objective functions \cite{Saaty2004Decision}.

    \item The performance of the proposed optimization
framework is evaluated on a real LV distribution network. Here, we do a comparison between our proposed optimization
framework, and the models that arbitrarily choose the CES
planning aspects such as its location, to assess the impact
of it on the objectives. Finally, a comprehensive analysis
of the results is also presented.
\end{itemize}

The rest of the paper is structured as follows. Section II presents the mathematical models used in our problem. The proposed CES planning and scheduling optimization framework is illustrated in Section III. Section IV is about the numerical and graphical results along with their discussion. Eventually, the conclusion of the work and possible future developments are given in Section V.

\section{{\normalsize System Mathematical Modelling}}

In this paper, the positive power absorption convention is considered for all nodes. Also, it is considered that there are multiple customers at each node. All the real and reactive power quantities are measured in kW and kVAR, respectively. It is assumed that power consumption (both real and reactive) and SPV generation of each customer are known ahead from their forecasts. A summary of the notations used in this paper, together with their definitions are given in the Nomenclature. 
	
\subsection{{\normalsize Power Flow Model}}
	
 \begin{figure}[t]
		\centering
			\includegraphics[width=2.7in,height=5.5cm]{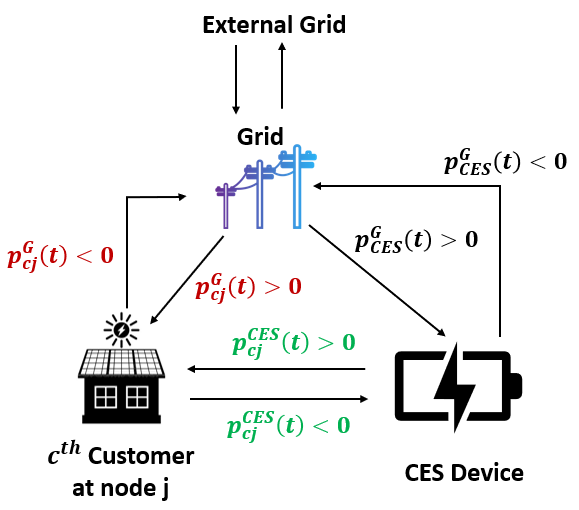}
		\caption{Possible power exchanges between a customer, the CES and the grid }\label{Fig:0fig2}
		\label{fig:Fig. 1}
\end{figure}

The typical mutual power exchanges that can ensue between different entities (i.e. customers, CES and grid) in the presence of a CES is shown in Fig. \ref{fig:Fig. 1}. $p_{cj}^{CES}(t)>0$ suggests a power import by the customer $c$ at node $j$ at time $t$ from the CES, and $p_{cj}^{CES}(t)<0$ occurs when that customer exports power to the CES. The same sign convention used for $p_{cj}^{CES}(t)$ is valid for $p_{cj}^{G}(t)$ and $p_{CES}^{G}(t)$. The mathematical relationships between the power flows shown in Fig. \ref{fig:Fig. 1} are described later.
	
The relationship between the line power flows and node absorptions which follows the LinDistflow model are given by (\ref{eq:1}), (\ref{eq:2}) \cite{Lin2018Decentralized}.

	\begingroup
    \setlength\abovedisplayskip{0pt}
        \begin{equation}\label{eq:1}
		P_{ij}(t)=p_{j}(t)+\sum_{k:j\rightarrow k}^{}P_{jk}(t)
	    \end{equation}
	    \begin{equation}\label{eq:2}
		Q_{ij}(t)=q_{j}(t)+\sum_{k:j\rightarrow k}^{}Q_{jk}(t)
	    \end{equation}
    \endgroup

The nodal real and reactive power absorptions are illustrated by the equations (\ref{eq:3}) and (\ref{eq:4}). The equation (\ref{eq:3a}) governs the real power absorption for the CES connected node and for the rest of the nodes (except the slack node), it is the equation (\ref{eq:3b}). Additionally, we assume all the SPV units and the CES operate at unity power factor.
	
	\begingroup
    \setlength\abovedisplayskip{0pt}
        \begin{subequations}\label{eq:3}
		\begin{equation}\label{eq:3a}
				\begin{split}
					p_{j}(t) =\sum_{c \in C_{j}}^{}p_{cj}^{L}(t)-\sum_{c \in C_{j}}^{}p_{cj}^{PV}(t)+ p_{j}^{CES,ch}(t) \\ - p_{j}^{CES,dis}(t) \hspace{3mm} \forall j\in \mathcal{V\setminus}\left \{ 0 \right \},  t \mathcal{\in T} 	
				\end{split}		
		\end{equation}

		\begin{equation}\label{eq:3b}
			\begin{split}
				p_{j}(t) =\sum_{c \in C_{j}}^{}p_{cj}^{L}(t)-\sum_{c \in C_{j}}^{}p_{cj}^{PV}(t) \hspace{3mm} \forall j\in \mathcal{V\setminus}\left \{ 0 \right \},  t \mathcal{\in T}	
			\end{split}		
		\end{equation}
	    \end{subequations}
	    
	    \begin{equation}\label{eq:4}
		\begin{split}
			q_{j}(t) =\sum_{c \in C_{j}}^{}q_{cj}^{L}(t) \hspace{3mm} \forall j\in \mathcal{V\setminus}\left \{ 0 \right \},  t \mathcal{\in T}  \\	 
		\end{split}
	    \end{equation}
  
    \endgroup

The equations (\ref{eq:5}) and (\ref{eq:6}) demonstrate how a customer exchanges power with the CES and the grid, when that customer encounters a mismatch of its real power consumption and SPV generation. When a customer experiences a deficit of its SPV generation to supply its real power consumption, that deficit can be fulfilled in share by the CES and the grid. On the other hand, if a customer has a surplus SPV generation, that customer exports the excess to both CES and the grid. The mathematical relationship between $p_{CES}^G(t)$, $p_{cj}^{CES}(t)$, $p_{j}^{CES,ch}(t)$ and $ p_{j}^{CES,dis}(t)$ can be written as (\ref{eq:7}).

	If $p_{cj}^{L}(t) \geq  p_{cj}^{PV}(t)$:
	
	\begingroup
    \setlength\abovedisplayskip{0pt}
        \begin{subequations}\label{eq:5}
		\begin{equation}
			0 \leq p_{cj}^{G}(t)+ p_{cj}^{CES}(t) = p_{cj}^{L}(t)-p_{cj}^{PV}(t) 
		\end{equation}
		
		\begin{equation}
			\begin{split}
				0 \leq p_{cj}^{G}(t) \leq  p_{cj}^{L}(t)-p_{cj}^{PV}(t) \\ \hspace{3mm} \forall j\in \mathcal{V\setminus}\left \{ 0 \right \},  c \in C_{j},  t \mathcal{\in T}
			\end{split}
		\end{equation}
	\end{subequations}
  
    \endgroup
	
	Otherwise:
	
	\begingroup
    \setlength\abovedisplayskip{0pt}
    \begin{subequations}\label{eq:6}
		\begin{equation}
			p_{cj}^{G}(t)+ p_{cj}^{CES}(t) = p_{cj}^{L}(t)-p_{cj}^{PV}(t) \leq 0
		\end{equation}
		
		\begin{equation}
			\begin{split}
				p_{cj}^{L}(t)-p_{cj}^{PV}(t) \leq p_{cj}^{G}(t) \leq 0 \\ \hspace{3mm} \forall j\in \mathcal{V\setminus}\left \{ 0 \right \},   c \in C_{j},  t \mathcal{\in T}  \\	
			\end{split}
		\end{equation}
	\end{subequations}

	\begin{equation}\label{eq:7}
		\begin{split}
			p_{CES}^G(t)= & \sum_{j=1}^{N}\left \{ \sum_{c \in C_{j}}p_{cj}^{CES}(t)+p_{j}^{CES,ch}(t) -p_{j}^{CES,dis}(t) \right \}
		\end{split}
	\end{equation}
	
	\endgroup

	The Lindistflow equations given in (\ref{eq:1})-(\ref{eq:4}) can be written in matrix format as ((\ref{eq:8}) \cite{Lin2018Decentralized}
	
	\begin{equation}\label{eq:8}
		\mathbf{U}=U_{0}\mathbf{1}-2\mathbf{\tilde{R}}\mathbf{p}-2\mathbf{\tilde{X}}\mathbf{q} \hspace{4mm} \forall t \mathcal{\in T}
	\end{equation}
	
where $\textbf{U} \! = \! |\textbf{V}(t)|^2$ is the vector of squared voltage magnitudes of nodes, $\mathbf{1}$ is a vector of all ones and $U_0 = \vert V_0 \vert^2$ is the squared voltage magnitude of the slack node. Also, $\textbf{p}$ and $\textbf{q}$ are the vectors of nodal real and reactive power absorption. The matrices $\mathbf{\tilde{R}}$ and  $\mathbf{\tilde{X}}$ $\in \mathbb{R}^{N\times N}$ have the elements $ {R}_{ij}=\sum{(a.b)\in L_{i}\cap L_{j}  }^{}r_{ab}$ and ${X}_{ij}=\sum{(a.b)\in L_{i}\cap L_{j}  }^{}x_{ab}$, respectively where $L_{i}$ is the set of lines on the path connecting node 0 and $``i"$ \cite{Chathurika2021Network,Lin2018Decentralized}.
	
The squared voltage magnitudes at each node needs to be maintained within its allowable voltage magnitude limits. This is guaranteed by the inequality given in (\ref{eq:9}). Here, $\mathbf{U_{min}}=U_{min}\mathbf{1}$ and $\mathbf{U_{max}}=U_{max}\mathbf{1}$.
	
	\begin{equation}\label{eq:9}
		\begin{split}
			\mathbf{U_{min}}\le \mathbf{U}\le \mathbf{U_{max}} \hspace{4mm} \forall t \mathcal{\in T}
		\end{split}  	
	\end{equation}
	
	\subsection{{\normalsize Community Energy Storage Model}}
In this section we present the mathematical modelling of the CES. We consider the CES is owned by a third party, and the owner is designated as the CES provider.

The set of constraints listed from (\ref{eq:10}) to (\ref{eq:17}) model the CES. The equations (\ref{eq:10}) and (\ref{eq:11}) imply that the CES charging and discharging power should not exceed the rated power $p_{j}^{Rate}$ of the CES. The temporal variation of the energy level of the CES is expressed by (\ref{eq:12}). Also, the CES energy level at any time should exist within its upper and lower state of charge (SoC) limits. This is handled by (\ref{eq:13}). The continuity of the CES operation over the next day is guaranteed by the inequality given in (\ref{eq:14}) which is bounded by a small positive number $\varepsilon$ \cite{Chathurika2021Network,Chathurika2020Community}. Note that $t_{d}$ in (\ref{eq:14}) represents the day number of the year. Here $t_{d} \in \mathcal{T}_{D}$, where $\mathcal{T}_{D}=\left \{ 1,2,....,N_{T}/24\right \}$ and $N_{T}$ is the cardinality of set $\mathcal T$.
	
	\begingroup
    \setlength\abovedisplayskip{0pt}
    {\normalsize \begin{equation}\label{eq:10}
			\begin{split}
				0 \le  p_{j}^{CES,ch}(t) \le p_{j}^{Rate} \hspace{3mm} \forall j\in \mathcal{V\setminus}\left \{ 0 \right \}, t \mathcal{\in T} 
			\end{split}
	\end{equation}}
	
	{\normalsize \begin{equation}\label{eq:11}
			\begin{split}
				0 \le  p_{j}^{CES,dis}(t) \le p_{j}^{Rate} \hspace{3mm} \forall j\in \mathcal{V\setminus}\left \{ 0 \right \}, t \mathcal{\in T} 
			\end{split}
	\end{equation}}
	
	{\normalsize \begin{equation}\label{eq:12}
			\begin{split}
				E_{j}^{CES}(t) & =E_{j}^{CES}(t-1)+(\eta^{ch}p_{j}^{CES,ch}(t) \\ & \hspace{4mm} -\frac{1}{\eta^{dis}}p_{j}^{CES,dis}(t)) \Delta t  \hspace{3mm} \forall j\in \mathcal{V\setminus}\left \{ 0 \right \}, \hspace{0.5mm} t \mathcal{\in T}
			\end{split}
		\end{equation}
	}
	{\normalsize \begin{equation}\label{eq:13}
			\lambda _{min} E_{j}^{cap} \le E_{j}^{CES}(t)\le \lambda _{max} E_{j}^{cap} \hspace{3mm} \forall j\in \mathcal{V\setminus}\left \{ 0 \right \}, t \mathcal{\in T}
		\end{equation}
	}
	{\normalsize \begin{equation}\label{eq:14}
			\begin{split}
				\left |  E_{j}^{CES}(24 t_{d})-E_{j}^{CES}(0) \right |\leq \varepsilon \hspace{3mm} \forall j\in \mathcal{V\setminus}\left \{ 0 \right \},  t_{d} \mathcal{\in T}_{D}
			\end{split}
	\end{equation}}

The equation (\ref{eq:15}) is used to find the optimal CES location. Also, (\ref{eq:15}) ensures that only one CES is installed in the network. If $a_{j}=0$, it implies that there is no CES at node $j$. If $a_{j}=1$, then the CES is connected to node $j$. To determine the optimal CES capacity $E_{j}^{cap}$, the inequality given in (\ref{eq:16}) is utilized. For a case $a_{j}=0$, (\ref{eq:16}) makes $E_{j}^{cap}$ also to be zero. When $E_{j}^{cap}=0$, the values $E_{j}^{CES}(t)$, $p_{j}^{CES,ch}(t)$ and $p_{j}^{CES,dis}(t)$ in (\ref{eq:12}) and (\ref{eq:13}) also turn out to be zero. The inequality in (\ref{eq:17}) guarantees the rated power of the CES is bounded by its minimum and maximum allowable values.

	{\normalsize \begin{equation}\label{eq:15}
			\begin{split}
				\sum_{j=1}^{N}a_{j}=1  \hspace{5mm} \forall j\in \mathcal{V\setminus}\left \{ 0 \right \},  a_{j} \in \left \{ 0,1 \right \} 
			\end{split}
	\end{equation}}
	
	{\normalsize \begin{equation}\label{eq:16}
			\begin{split}
				a_{j} E_{min}^{cap}  \le  E_{j}^{cap} \le a_{j} E_{max}^{cap}  \hspace{4mm} \forall j\in \mathcal{V\setminus}\left \{ 0 \right \}, a_{j} \in \left \{ 0,1 \right \} 
			\end{split}
	\end{equation}}
	
	{\normalsize \begin{equation}\label{eq:17}
			\begin{split}
				a_{j} p_{min}^{Rate} \le  p_{j}^{Rate} \le a_{j} p_{max}^{Rate}   \hspace{4mm} \forall j\in \mathcal{V\setminus}\left \{ 0 \right \}, a_{j} \in \left \{ 0,1 \right \} 
			\end{split}
	\end{equation}}
  
    \endgroup
	
	\section{{\normalsize Optimization Framework \& Problem Formulation}}

In our paper, it is expected to minimize the real power loss of the network, energy trading costs of the customers and the CES provider with the grid, and to minimize the CES investment cost. Therefore, a multi-objective function is obtained by combining those objectives functions, and its formulation is given as follows.

	\subsection{{\normalsize Objective Functions}}
	\subsubsection{{\normalsize Minimizing the Real Power Loss of the Network}}
	The real power loss in a network can be written as (\ref{eq:19}), in terms of (\ref{eq:18}), and by taking $U_{i}(t)\approx U_{0}(t) \hspace{2mm} \forall i\in \mathcal{V\setminus}\left \{ 0 \right \}$ \cite{Chathurika2020Community,Lin2018Decentralized}.

    \begin{equation}\label{eq:18}
		\left| I_{ij}(t) \right|^{2}=\frac{P_{ij}(t)^2+Q_{ij}^2(t)}{U_i(t)} \hspace{4mm} \forall (i,j) \mathcal{\in E}, t \mathcal{\in T}
	\end{equation}
	
\begin{equation}\label{eq:19}
f_{Ploss}=\sum_{t \mathcal{ \in T}}\sum_{(i,j)\in \mathcal{E} }^{}r_{ij}\left| I_{ij}(t) \right|^{2} 
\end{equation}

\subsubsection{{\normalsize Minimizing the Energy Trading Cost of the Customers and the CES Provider with the Grid}}

The first term of the objective function given in (\ref{eq:20}) relates to the energy trading cost with the grid by customers, and latter for the CES provider.

\begin{equation}\label{eq:20}
	f_{En,cost}= \sum_{t \mathcal{ \in T}}\lambda _{p}(t) \left \{ \sum_{j=1}^{N}\sum_{c \in C_{j}}p_{cj}^{G}(t)+p_{CES}^G(t) \right \}\Delta t
\end{equation}

Here, it is considered a one-for-one non-dispatchable energy buyback scheme  such that the same energy price for both imports and exports of energy from the grid by the customers and the CES is used \cite{Martin1to1}. This kind of an energy pricing scheme can effectively value the SPV power as being same as the power imported from the grid, which is usually generated by a conventional generation method.

\subsubsection{{\normalsize Minimizing the Investment Cost of the CES}}

The third objective is to minimize the investment cost of the CES device which is given by (\ref{eq:21}) \cite{Zheng2020Hierarchical}.

\begin{equation}\label{eq:21}
	f_{Inv,cost}= \gamma_{CES}+\delta_{CES}E_{j}^{cap}
\end{equation}

\subsection{{\normalsize Problem Formulation}}

The three objective functions are normalized and weighted to form the multi-objective function in (\ref{eq:22}), according to the techniques described in \cite{Grodzevich2006Normalization}. The normalization guarantees the objective functions are converted into a form which can be added together (since $f_{Ploss}$ is measured in kW, and $f_{En,cost},f_{Inv,cost}$ are measured in AUD ). 

\begin{equation}\label{eq:22}
	\begin{split}
		\resizebox{1.02\hsize}{!}{$\mathbf{x^{*}}=\underset{\mathbf{x} \in X}{argmin}\hspace{3mm} w_{1}\left \{ \frac{f_{Ploss}-f_{Ploss}^{utopia}}{f_{Ploss}^{Nadir}-f_{Ploss}^{utopia}} \right \} + w_{2}\left \{ \frac{f_{En,cost}-f_{En,cost}^{utopia}}{f_{En,cost}^{Nadir}-f_{En,cost}^{utopia}} \right \}$} \\ \resizebox{0.43\hsize}{!}{$+w_{3}\left \{ \frac{f_{Inv,cost}-f_{Inv,cost}^{utopia}}{f_{Inv,cost}^{Nadir}-f_{Inv,cost}^{utopia}} \right \}$} \hspace{30mm} 
	\end{split}
\end{equation}

where $X$ is the feasible set which is constrained by (\ref{eq:1})-(\ref{eq:17}). The utopia values, individual minimum point values and nadir values of the multi-objective function are found by (\ref{eq:23}), (\ref{eq:24}) and (\ref{eq:25}), respectively. Besides, the decision variable vector can be explicitly expressed as (\ref{eq:26}).

\begin{figure}[t]
	\centering
	\includegraphics[width=2.8in,height=4.3cm]{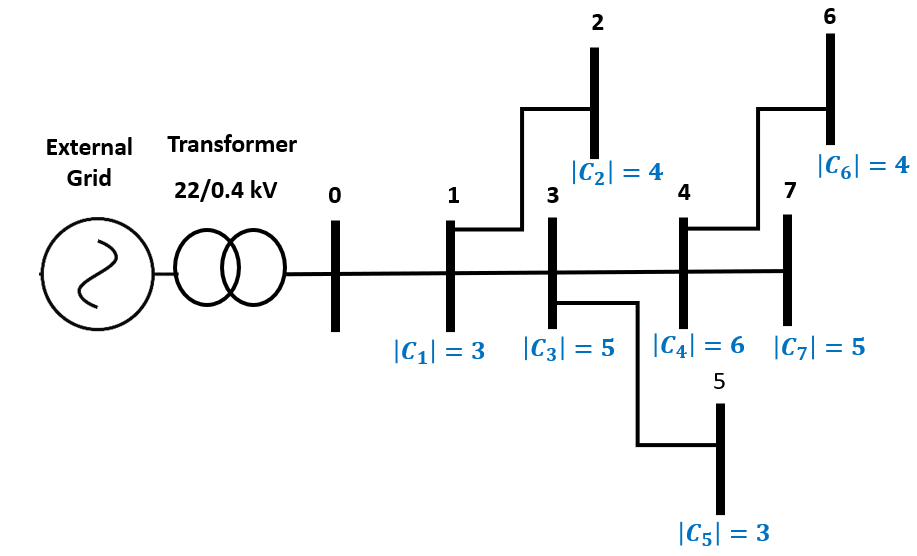}
	\caption{7-Node LV radial distribution network}\label{Fig:0fig3}
	\label{fig:Fig. 2}
	\vspace{-1em}
\end{figure}

\begingroup
    \setlength\abovedisplayskip{0pt}
    \begin{subequations}\label{eq:23}
	\begin{equation}\label{eq:23a}
		f_{Ploss}^{utopia}={f_{Ploss}(\mathbf{x^{\ast}_{Ploss}})}
	\end{equation}
	\begin{equation}\label{eq:23b}
		f_{En,cost}^{utopia}={f_{En,cost}(\mathbf{x^{\ast}_{En,cost}})}
	\end{equation}
	\begin{equation}\label{eq:23c}
		f_{Inv,cost}^{utopia}={f_{Inv,cost}(\mathbf{x^{\ast}_{Inv,cost}})}
	\end{equation}
\end{subequations}

\begin{subequations}\label{eq:24}
	\begin{equation}\label{eq:24a}
	\mathbf{x^{\ast}_{Ploss}}=\underset{\mathbf{x} \in X}{argmin}\hspace{3mm} f_{Ploss}
	\end{equation}
	\begin{equation}\label{eq:24b}
		\mathbf{x^{\ast}_{En,cost}}=\underset{\mathbf{x} \in X}{argmin}\hspace{3mm} f_{En,cost}
	\end{equation}
	\begin{equation}\label{eq:24c}
		\mathbf{x^{\ast}_{Inv,cost}}=\underset{\mathbf{x} \in X}{argmin}\hspace{3mm} f_{Inv,cost}
	\end{equation}
\end{subequations}

\begin{subequations}\label{eq:25}
	\begin{equation}\label{eq:25a}
         \begin{split}
            f_{Ploss}^{Nadir}= Max  \bigl\{  f_{Ploss}(\mathbf{x^{\ast}_{Ploss}}), f_{Ploss}(\mathbf{x^{\ast}_{En,cost}}) , \\ f_{Ploss}(\mathbf{x^{\ast}_{Inv,cost}}) \bigl \}
        \end{split}
	\end{equation}
	\begin{equation}\label{eq:25b}
 \begin{split}
            f_{En,cost}^{Nadir}= Max  \bigl\{  f_{En,cost}(\mathbf{x^{\ast}_{Ploss}}), f_{En,cost}(\mathbf{x^{\ast}_{En,cost}}) , \\ f_{En,cost}(\mathbf{x^{\ast}_{Inv,cost}}) \bigl \}
        \end{split}
	\end{equation}
	\begin{equation}\label{eq:25c}
        \begin{split}
            f_{Inv,cost}^{Nadir}= Max  \bigl\{  f_{Inv,cost}(\mathbf{x^{\ast}_{Ploss}}), f_{Inv,cost}(\mathbf{x^{\ast}_{En,cost}}) , \\ f_{Inv,cost}(\mathbf{x^{\ast}_{Inv,cost}}) \bigl \}
        \end{split}
	\end{equation}
\end{subequations}

\begin{equation}\label{eq:26}
    \mathbf{x=(a_{j}, p_{j}^{Rate},E_{j}^{cap}, p_{j}^{CES,ch}, p_{j}^{CES,dis}, p_{CES}^G,p_{cj}^{G})}
\end{equation}
\endgroup

In summary, the optimization framework can be written as (\ref{eq:22}), subject to a set of constraints (\ref{eq:1})-(\ref{eq:17}). Also, as (\ref{eq:22}) being a quadratically-constrained convex multi-objective function, it is solved as a mixed-integer quadratic program.

\section{{\normalsize Numerical and Simulation Results}}

In the simulations, a 7-node LV radial distribution network given in Fig. \ref{fig:Fig. 2} is used and its line data can be found in \cite{Zeraati2018Distributed}. Also, real power consumption and SPV generation data of 30 customers in an Australian residential community  were used for simulations \cite{AusgridData}. To be more practical, we randomly allocated multiple customers for each node. Hence, $\sum_{j=1}^{N}\left | C_{j} \right |=30$, and the number of customers at each node are marked in Fig. \ref{fig:Fig. 2}. Here, all the customers generate SPV power in addition to their real power consumption. However, reactive power consumption of the customers is not considered due to the lack of sufficient real data. As the optimization involves not only a scheduling problem but also a planning problem, the optimization is performed over a long time period. Thus, we consider one year time period split in one hour time intervals (i.e.$\left | \mathcal T \right |=8760$ ) for the simulations.

The voltage and power base are taken as 400V and 100 kVA, respectively. In addition to that, $V_{0}=1 p.u.$, $U_{min}=0.9025 p.u.$, $U_{max}=1.1025 p.u.$, $\lambda _{min}=0.05$, $\lambda _{max}=1$, $\eta^{ch}=0.98$, $\eta^{dis}=1.02 $, $E_{min}^{cap}=200 kWh$, $E_{max}^{cap}=2000 kWh$, $p_{min}^{Rate}=20 kW$, $p_{max}^{Rate}=200 kW$, $\varepsilon=0.0001 kWh $  and $\Delta t= 1h$  are used as the model parameters. The values of $\gamma_{CES}$ and $\delta_{CES}$ are taken as 24000 AUD and 300 AUD/kWh as specified in \cite{Zheng2020Hierarchical}. Additionally, the weighting factors $w_{1}, w_{2}$ and $w_{3}$ were calculated according to the principles of AHP specified in \cite{Saaty2004Decision}. We considered a moderate plus importance for both  $f_{En,cost}$ and  $f_{Inv,cost}$ compared to $f_{Ploss}$, and an equal importance for $f_{En,cost}$ and $f_{Inv,cost}$. Hence, based on the AHP method, the values of $w_{1}, w_{2}$ and $w_{3}$ were calculated as 1/9, 4/9 and 4/9, respectively. Fig. \ref{fig:Fig. 3} depicts how the grid energy price varies with the time of the day following a time of use (ToU) tariff scheme. As seen in Fig. \ref{fig:Fig. 3}, the grid energy price is 0.24871 AUD/kWh during $T_{1}$(from 12am-7am) \& $T_{5}$(from 10pm-12am), 0.31207 AUD/kWh during $T_{2}$(from 7am-3pm) \& $T_{4}$ (from 9pm-10pm) and 0.52602 AUD/kWh during $T_{3}$(from 3pm-9pm) \cite{VICToU}. 

\begin{figure}[t]
	\centering
	\includegraphics[width=3.4in,height=4.3cm]{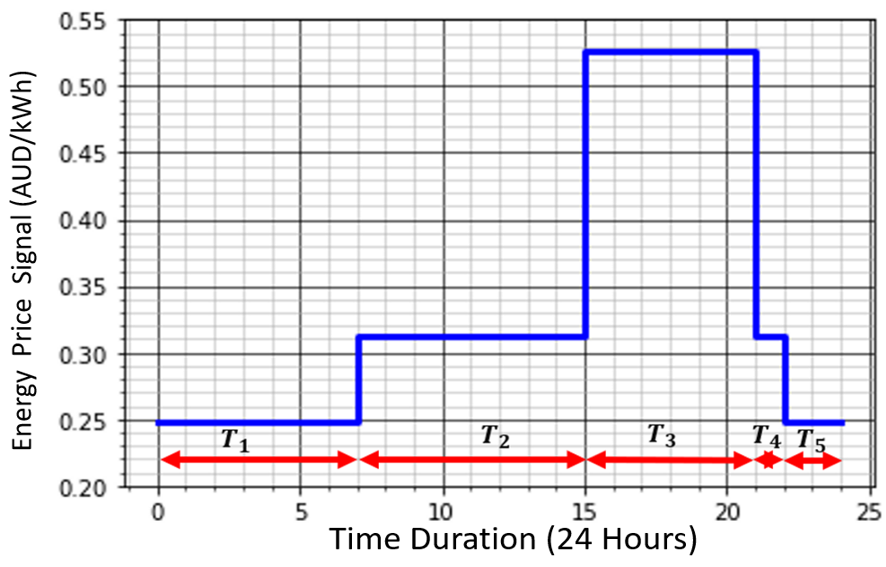}
	\caption{Variation of grid energy price for 24 hours}\label{Fig:0fig9}
	\label{fig:Fig. 3}
	\vspace{-1em}
\end{figure}

\begin{table*}[]
	\centering
	\captionsetup{justification=centering}
	\caption{SUMMARY OF THE RESULTS FOR CASE STUDIES }
	\label{table:1}
	\renewcommand{\arraystretch}{1.3}
	\begin{tabular}{|c|c|c|c|c|c|c|}
		\hline
		& \begin{tabular}[c]{@{}c@{}}CES\\ Location\\ (Node)\end{tabular} & \begin{tabular}[c]{@{}c@{}}Optimal CES\\ Capacity (kWh)\end{tabular} & \begin{tabular}[c]{@{}c@{}}Optimal CES\\ Power Rating (kW)\end{tabular} & \begin{tabular}[c]{@{}c@{}}Real Energy\\ Loss$^1$ (kWh)\end{tabular} & \begin{tabular}[c]{@{}c@{}}Energy Trading\\ Cost With \\ Grid$^1$ (AUD)\end{tabular} & \begin{tabular}[c]{@{}c@{}}CES Investment\\ Cost (AUD)\end{tabular} \\ \hline
		\begin{tabular}[c]{@{}c@{}}Base Case (Without CES)\end{tabular} & Not applicable                                                  & Not applicable                                                       & Not applicable                                                            & 110116.68                                                        & 45585                                                               & Not applicable                                                      \\ \hline
		\begin{tabular}[c]{@{}c@{}}Case I (Proposed Model)\end{tabular}                                                      & 4 (optimal)                                                     & 482.15                                                               & 200                                                                       & 78200.88 (71.02\%)                                                         & 38520 (84.50\%)                                                               & 168645                                                              \\ \hline
		Case II                                                           & 3 (chosen)                                                      & 601.32                                                               & 200                                                                       & 80250.48 (72.88\%)                                                         & 43362  (95.12\%)                                                             & 204396                                                              \\ \hline
		Case III                                                          & 5 (chosen)                                                      & 601.32                                                               & 200                                                                       & 81961.28 (74.43\%)                                                          & 43840 (96.17\%)                                                              & 204396                                                              \\ \hline
		Case IV                                                           & 6 (chosen)                                                      & 482.15                                                               & 200                                                                       & 80761.16 (73.34\%)                                                         & 44154 (96.86\%)                                                              & 168645                                                              \\ \hline
		Case V                                                            & 7 (chosen)                                                      & 547.69                                                               & 200                                                                       & 86082.52 (78.17\%)                                                          & 43625 (95.70\%)                                                               & 188307                                                              \\ \hline
		\multicolumn{7}{l}{$^1$ Percentage values are calculated with respect to their corresponding values without a CES}  
	\end{tabular}
\end{table*}

\subsection{{\normalsize Case Study - Proposed Optimization Framework Vs Optimization Models With Arbitrary CES Locations}}

We did a case study to compare the results of our model with four different cases by arbitrarily changing the CES location. For this, we considered our optimization framework as Case I, while the rest as Case II-V. The same optimization framework (except the constraint that finds the optimal CES location), and the model parameters as for Case I were used for Case II-V. A synopsis of the results for the five cases are tabulated in Table \ref{table:1}. The Case I lists the planning results and the minimized objective function values for our proposed model. The Cases II and III suggest the same optimal CES capacity and the rated power. Nevertheless, due to their difference in CES location, Case II provides less real energy loss and energy trading cost compared with the Case III. When the CES is at node 6, the optimization suggests the same optimal capacity as in Case I. However, as node 6 is not the optimal location for CES, the real energy loss and energy trading cost for Case IV are higher than in Case I. Also, our model has produced the highest cost reduction percentages for real energy loss (28.98\%) and the energy trading cost with the grid (15.5\%), compared to all the other cases. Hence, it is clear that Case I yields the minimum values for all the three objective functions, and this justifies the effectiveness of our optimization framework compared to the models that optimize only the CES scheduling.

\begin{figure}[t]
	\centering
	\includegraphics[width=3.5in,height=5.1cm]{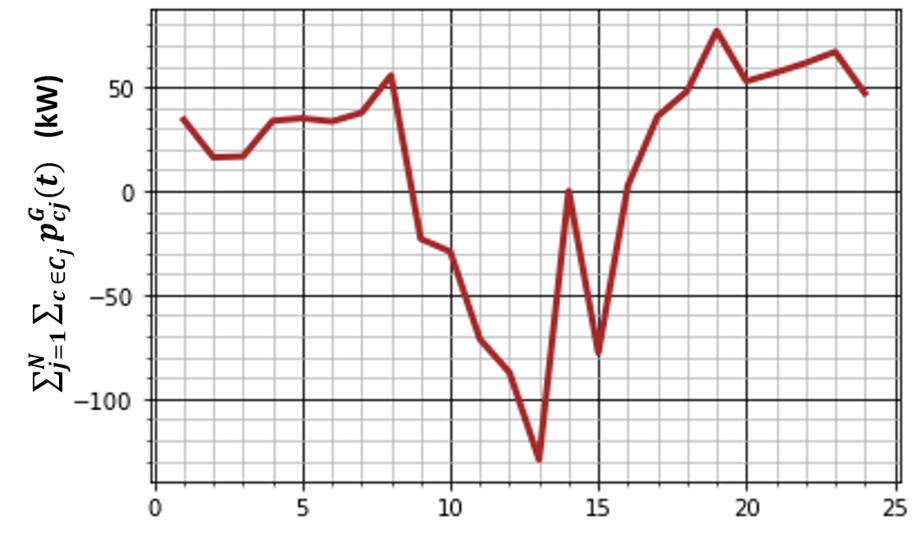}
	\caption{Total power exchange with the grid by the customers}\label{fig:Fig. 4}
\end{figure}

\begin{figure}[t]
	\centering
	\includegraphics[width=3.5in,height=5cm]{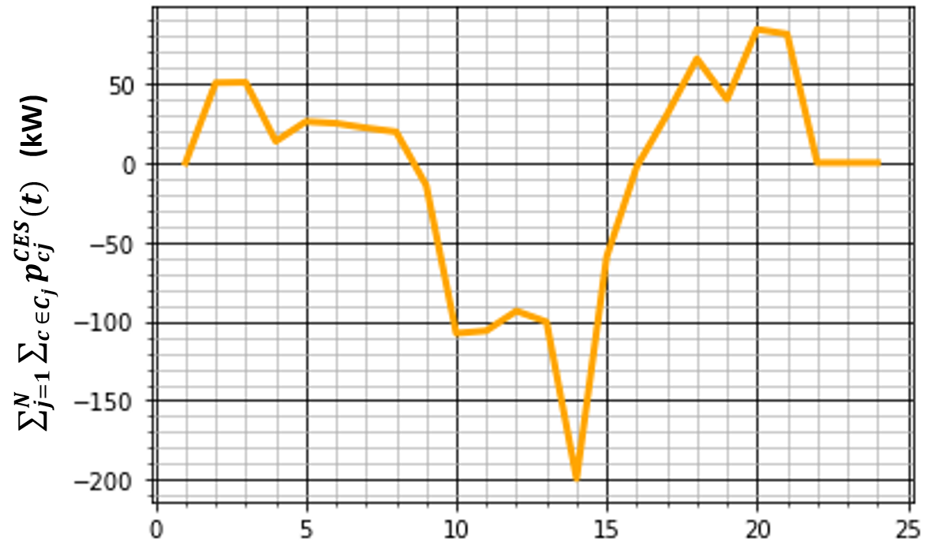}
	\caption{Total power exchange with the CES by the customers}\label{fig:Fig. 5}
\end{figure}

\begin{figure}[t]     
         \centering
         \includegraphics[width=3.4in,height=5cm]{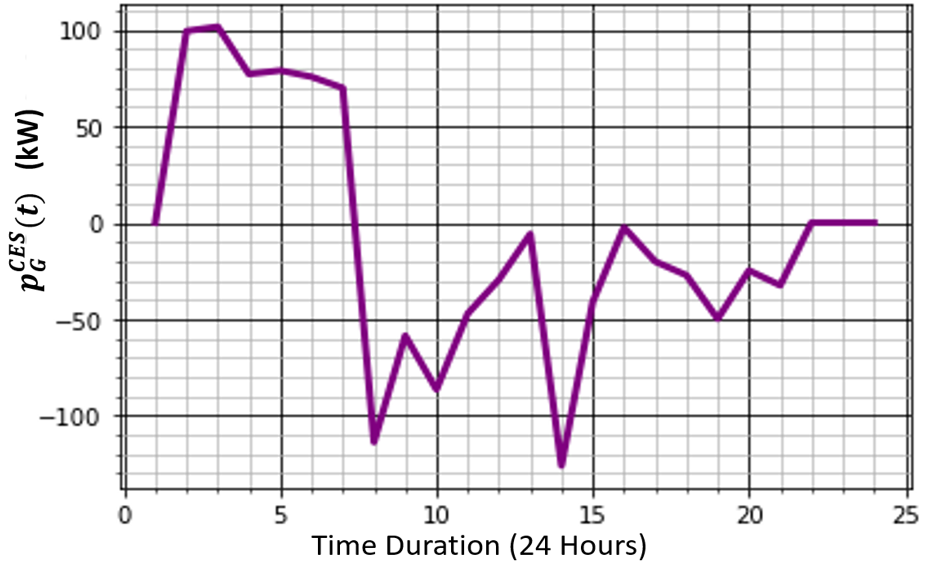}
         \caption{ Power exchange with the grid by the CES}\label{fig:Fig. 6}
\end{figure}

\subsection{{\normalsize Analysis of the Results-Mutual Power Exchanges Between the customers, the CES and the grid}}

In order to understand the CES scheduling and power exchanges between different entities, we select a single day (24 hours) for our discussion. Fig. \ref{fig:Fig. 4} shows the variation of total power exchange that occurs with the grid by the customers. Since $\sum_{j=1}^{N}\sum_{c \in C_{j}}^{} p_{cj}^{G}(t)$ being a positive value approximately during $T_{1}, T_{3}, T_{4}, T_{5}$ time intervals, it implies that the customers tend to import certain amount of power from the grid for satisfying their real power consumption during those time periods. On the other hand,  during $T_{2}$ (time period of the day usually the SPV generation is high), the customers have a tendency to export a portion of their surplus SPV generation to the grid. This is evident as $\sum_{j=1}^{N}\sum_{c \in C_{j}}^{} p_{cj}^{G}(t)<0$ during $T_{2}$. This behavior guarantees a cost benefit for the customers for their exported power according to equation (\ref{eq:20}).

\begin{figure}[t]
         \centering
         \includegraphics[width=3.4in,height=5.4cm]{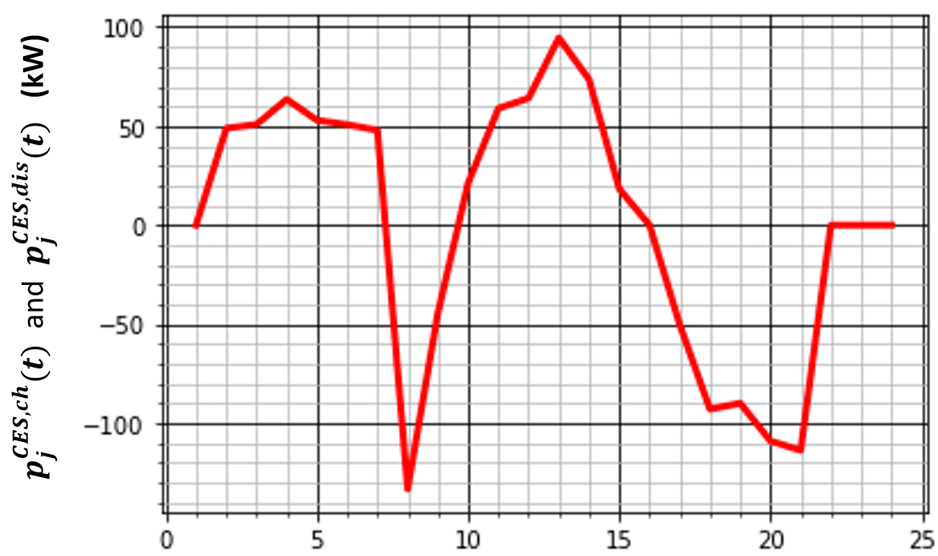}
         \caption{CES charging and discharging power pattern}\label{fig:Fig. 7}
\end{figure}

\begin{figure}[t]
         \centering
         \includegraphics[width=3.4in,height=5.4cm]{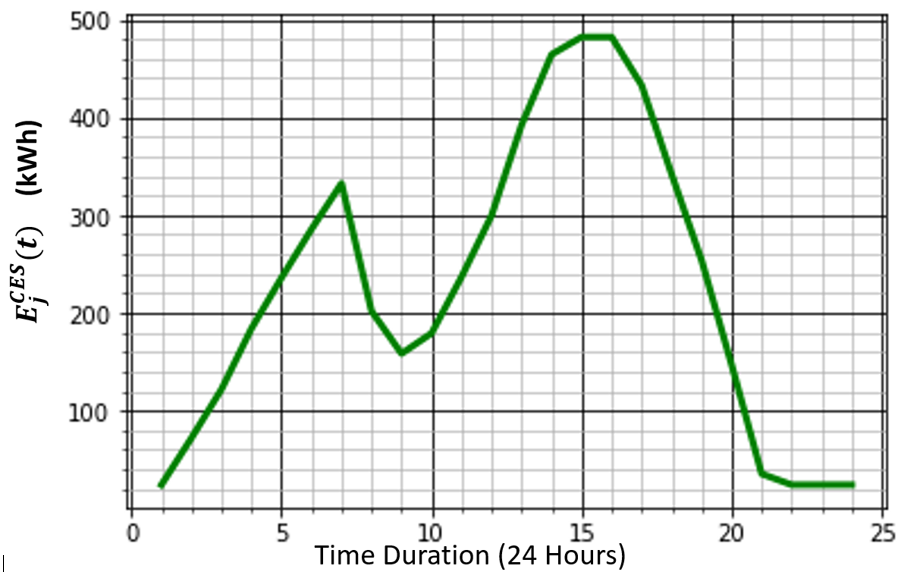}
      \caption{Temporal variation of the CES energy level}\label{fig:Fig. 8}
\end{figure}

The Fig. \ref{fig:Fig. 5} depicts how the customers exchange power with the CES. During $T_{2}$, the customers export a part of their surplus SPV generation to the CES. On the contrary, during rest of the time periods, the customers import a certain amount of power from the CES for satisfying their real power consumption. This action results in reducing the cost for the customers as the amount of power imported from the grid is minimized. 

The Fig. \ref{fig:Fig. 6} illustrates how the CES exchanges power with the grid. As the grid energy price during $T_{1}$ being the lowest, the CES tends to import power from the grid (i.e. $p_{G}^{CES}(t) > 0$) during $T_{1}$. This guarantees that the CES is charged with low priced energy from the grid. However, during $T_{2}$, $T_{3}$ and $T_{4}$,  it is seen that the CES exports its power back to the grid (i.e. $p_{G}^{CES}(t) < 0$). This happens as the CES provider can maximize its revenue by exporting power back to grid.

In Fig. \ref{fig:Fig. 7} and \ref{fig:Fig. 8}, it is observed that during T1, the CES charges (from the low priced grid energy) and partially discharges by the end of T1. During T2, the CES continues to charge and by the end of this time period, it reaches its maximum energy level. The stored energy in the CES is fully utilized during T3 and T4 for partially supplying the real power consumption of the customers. This facilitates monetary benefits for both the customers as the amount of expensive power imported from the grid is lowered. Additionally, when observing the temporal variation of the CES energy level, it is visualized that it is the peak value of the CES energy level which was obtained as the optimal CES capacity (i.e. 482.15 kWh).

\section{{\normalsize Conclusion \& Future Work}}

In this work, we have explored how the optimization of the 
planning and scheduling aspects of a community energy storage (CES) can benefit both the network and the customers. To this end, we developed a multi-objective mixed-integer quadratic optimization framework to minimize three objectives: (i) network real power
loss, (ii) energy trading cost of the customers and the CES provider with the grid, and (iii) the CES investment cost. The simulation results highlighted our optimization framework is competent in acquiring the expected merits compared with the case without
a CES, and optimization models that optimize only the scheduling of CES.

As future work, we expect to develop the work considering a stochastic model taking into account the uncertainties of real power consumption and SPV generation of the customers. Moreover, we look forward to extend the work by considering the unbalanced nature of LV distribution networks, and reactive power control capabilities of solar photovoltaic (SPV) and CES inverters.

\end{document}